# On the Separability of Parallel Gaussian Interference Channels

Sang Won Choi
School of EECS
KAIST
Daejeon, Korea
Email: ace1905@kaist.ac.kr

Sae-Young Chung
School of EECS
KAIST
Daejeon, Korea
Email: sychung@ee.kaist.ac.kr

*Abstract*—The separability in parallel Gaussian interference channels (PGICs) is studied in this paper. We generalize the separability results in one-sided PGICs (OPGICs) by Sung *et al.* to two-sided PGICs (TPGICs). Specifically, for strong and mixed TPGICs, we show necessary and sufficient conditions for the separability. For this, we show diagonal covariance matrices are sum-rate optimal for strong and mixed TPGICs.

## I. INTRODUCTION

Interference is a fundamental problem in wireless communications. Studying the interference channel (IC) can give us insights into how to deal with the problem. Specifically, 2-user single-input single-output (SISO) Gaussian IC (GIC) [1]-[6] has been studied. The capacity regions of the 2-user GICs have been clarified for very strong [1] and strong [2], [3] interference cases. Recently, the sum capacity for the very weak interference has been discovered [4]-[6], where a proper noisy version of genie is used for the tight upper bound on the sum capacity.

So far, most of the research on the IC has been focussed on the GIC itself. Recently, parallel GIC (PGIC) has been studied with interest. Here, the PGIC has several independent GICs as sub-channels. In the PGIC, joint coding and independent coding can be considered, where joint coding means coding over multiple sub-channels, and independent coding refers to coding over each sub-channel separately.

In 2-user PGICs, there have been investigations into whether independent coding suffices to achieve the (sum) capacity, i.e., whether the separability holds or not. In [7], the separability has been considered for one-sided PGIC (OPGIC), which has been further studied in the ergodic sense in [8]. Recently, the independent coding has been shown to achieve the sum capacity in the noisy interference regime [9], where treating interference as noise in each sub-channel is optimal in the sense of the sum capacity.

The main contribution of this paper is in considering the separability in a class of strong, mixed, and weak PGICs. This paper is organized as follows. In Section II, the channel model is described. In Section III, we show our separability results in the sense of the sum capacity. Then, we conclude the paper in Section IV.

## II. CHANNEL MODEL

We consider PGICs described as

$$\mathbf{Y}_1 = \mathbf{H}_{11}\mathbf{X}_1 + \mathbf{H}_{21}\mathbf{X}_2 + \mathbf{N}_1 \tag{1}$$

and

$$\mathbf{Y}_2 = \mathbf{H}_{12}\mathbf{X}_1 + \mathbf{H}_{22}\mathbf{X}_2 + \mathbf{N}_2, \tag{2}$$

where $\mathbf{Y}_k = [y_{k1}\ y_{k2}\ \cdots\ y_{kM}]^T \in \mathbb{R}^M$, $\mathbf{X}_k = [x_{k1}\ x_{k2}\ \cdots\ x_{kM}]^T \in \mathbb{R}^M$, $k = 1,\ 2$, where $M$ is the number of sub-channels, $y_{km}$ ($x_{km}$) is the received (transmitted) signal at the $k$-th receiver (transmitter) in the $m$-th sub-channel for $m = 1,\ 2,\ \cdots,\ M$, and

$$\mathbf{H}_{kl} = \begin{pmatrix} h_{kl,1} & 0 & \cdots & 0 \\ 0 & h_{kl,2} & \cdots & 0 \\ \vdots & \vdots & \ddots & \vdots \\ 0 & 0 & \cdots & h_{kl,M} \end{pmatrix}$$

is a diagonal channel matrix whose $(m, m)$-th component is the non-zero channel coefficient from the $k$-th transmitter to the $l$-th receiver in the $m$-th sub-channel. The noise vectors $\mathbf{N}_1$ and $\mathbf{N}_2$ are additive white Gaussian with zero mean and covariance matrix of $\mathbf{I}_M$. Here, $\mathbf{I}_M$ denotes the $M \times M$ identity matrix.

At the $k$-th transmitter, a message $M_k$ uniformly distributed over the message index set $\{1,\ 2,\ \cdots,\ 2^{nR_k}\}$ is mapped to the transmitted codeword $[\mathbf{X}_{k,1},\ \mathbf{X}_{k,2},\ \cdots,\ \mathbf{X}_{k,n}]$ of length $n$, where $\mathbf{X}_{k,i} = [x_{k1,i}\ x_{k2,i}\ \cdots x_{kM,i}]^T$ for $i = 1,\ 2,\ \cdots,\ n$, which is subject to an average power constraint per sub-channel, i.e.,

$$\frac{1}{n}\sum_{i=1}^{n} |x_{km,i}|^2 \le P_{km}. \tag{3}$$

At the $k$-th receiver, the block of $n$ received signal vector $[\mathbf{Y}_{k,1},\ \mathbf{Y}_{k,2},\ \cdots, \mathbf{Y}_{k,n}]$ is used to decode the message and an error happens when the output of the decoder $\hat{M}_k \ne M_k$.

The error probability for the $k$-th user is given by

$$\lambda_{k,n} = \Pr(\hat{M}_k \ne M_k) \tag{4}$$

assuming uniform distribution for messages. A rate pair $(R_1, R_2)$ is said to be achievable if we have a sequence of encoding and decoding functions such that $\lambda_n \to 0$ as $n \to \infty$, where $\lambda_n$ is the maximum of $\lambda_{1,n}$ and $\lambda_{2,n}$. The capacity region for the PGIC is defined as the closure of the set of all achievable rate pairs.

Note that the PGIC of (1) and (2) is a special form of the general multiple-input multiple-output (MIMO) GIC [10] with diagonal $\mathbf{H}_{kl}$'s for $k, l = 1, 2$.

## III. SUM CAPACITY

In this section, we analyze the sum capacity under joint coding and under independent coding. Then, we investigate whether the separability holds or not for some classes of TPGICs. We start by stating the following lemma.

*Lemma 1:* Let $\mathbf{w} = [1\ w_2\ \cdots\ w_K]^T$ be a weight vector whose $k$-th element is the weight for the $k$-user's rate with $w_k \geq 0$. A covariance matrix for the $k$-th user is denoted as $\mathbf{S}_k$ with constraint $\mathrm{diag}(\mathbf{S}_k) \leq \mathbf{P}_k$, where $\mathrm{diag}(\mathbf{S}_k)$ and $\mathbf{P}_k$ are diagonal matrices of size $M \times M$ whose $(m,m)$-th components are the $(m,m)$-th element of $\mathbf{S}_k$ and $P_{km}$, respectively. Then, the weighted sum capacity of MIMO IC denoted as $f(\mathbf{w}, \mathbf{P}_1, \mathbf{P}_2, \cdots, \mathbf{P}_K)$ is concave in $(\mathbf{P}_1, \mathbf{P}_2, \cdots, \mathbf{P}_K)$.

*Proof:* It is directly from [9]. Specifically, the set of all achievable schemes with $\mathrm{diag}(\mathbf{S}_k) \leq \mathbf{P}_k$ always includes TDM/FDM between any two achievable schemes with $\mathrm{diag}(\mathbf{S}_k) \leq \mathbf{P}'_k$ and $\mathrm{diag}(\mathbf{S}_k) \leq \mathbf{P}''_k$, where $\mathbf{P}_k = \lambda \mathbf{P}'_k + (1-\lambda)\mathbf{P}''_k$ for $0 \leq \lambda \leq 1$. ∎

### A. Capacity region for PGICs

*Notation 1:* For full-rank square matrices $\mathbf{S}$ and $\mathbf{T}$, $\mathbf{S} \geq \mathbf{T}$ ($\mathbf{S} > \mathbf{T}$) means that $\mathbf{S} - \mathbf{T}$ is positive semidefinite (definite). Let $\mathbf{H}^{[m]}$ denote $(h_{11,m}, h_{12,m}, h_{21,m}, h_{22,m})$ for $m = 1, 2, \cdots, M$. For notational convenience, we use the following notations:

$$A_m = \frac{1}{2}\log_2\left(1 + |h_{11,m}|^2 P_{1m}\right), \quad (5)$$

$$B_m = \frac{1}{2}\log_2\left(1 + |h_{12,m}|^2 P_{1m}\right), \quad (6)$$

$$C_m = \frac{1}{2}\log_2\left(1 + |h_{21,m}|^2 P_{2m}\right), \quad (7)$$

$$D_m = \frac{1}{2}\log_2\left(1 + |h_{22,m}|^2 P_{2m}\right), \quad (8)$$

$$E_m = \frac{1}{2}\log_2\left(1 + |h_{11,m}|^2 P_{1m} + |h_{21,m}|^2 P_{2m}\right), \quad (9)$$

$$F_m = \frac{1}{2}\log_2\left(1 + |h_{12,m}|^2 P_{1m} + |h_{22,m}|^2 P_{2m}\right), \quad (10)$$

$$G_m = \frac{1}{2}\log_2\left(1 + \frac{|h_{11,m}|^2 P_{1m}}{1 + |h_{21,m}|^2 P_{2m}}\right), \quad (11)$$

$$H_m = \frac{1}{2}\log_2\left(1 + \frac{|h_{22,m}|^2 P_{2m}}{1 + |h_{12,m}|^2 P_{1m}}\right), \quad (12)$$

$$I_m = \frac{1}{2}\log_2\left(1 + |h_{21,m}|^2 P_2 + \frac{|h_{11,m}|^2 P_1}{1 + |h_{12,m}|^2 P_1}\right), \quad (13)$$

and

$$J_m = \frac{1}{2}\log_2\left(1 + |h_{12,m}|^2 P_1 + \frac{|h_{22,m}|^2 P_2}{1 + |h_{21,m}|^2 P_2}\right) \quad (14)$$

for $m = 1, 2, \cdots, M$.

*1) Strong TPGIC:*

*Lemma 2:* For strong TPGIC, i.e., $\mathbf{H}_{12}^2 \geq \mathbf{H}_{11}^2$ and $\mathbf{H}_{21}^2 \geq \mathbf{H}_{22}^2$, the capacity region is given by

$$\bigcup_{\mathrm{diag}(\mathbf{S}_k) \leq \mathbf{P}_k,\ k=1,\ 2} \left\{ (R_1,\ R_2) \Big|
\begin{array}{l}
0 \leq R_1 \leq \frac{1}{2}\log_2 \left|\mathbf{I}_M + \mathbf{H}_{11}\mathbf{S}_1\mathbf{H}_{11}^T\right|, \\
0 \leq R_2 \leq \frac{1}{2}\log_2 \left|\mathbf{I}_M + \mathbf{H}_{22}\mathbf{S}_2\mathbf{H}_{22}^T\right|, \\
0 \leq R_1 + R_2 \leq \frac{1}{2}\log_2 \left|\mathbf{I}_M + \left(\mathbf{H}_{11}\mathbf{S}_1\mathbf{H}_{11}^T + \mathbf{H}_{21}\mathbf{S}_2\mathbf{H}_{21}^T\right)\right|, \\
0 \leq R_1 + R_2 \leq \frac{1}{2}\log_2 \left|\mathbf{I}_M + \left(\mathbf{H}_{12}\mathbf{S}_1\mathbf{H}_{12}^T + \mathbf{H}_{22}\mathbf{S}_2\mathbf{H}_{22}^T\right)\right|
\end{array} \right\}, \quad (15)$$

where the corresponding sum capacity is given by

$$\min\left(\sum_{m=1}^M A_m + D_m,\ \sum_{m=1}^M E_m,\ \sum_{m=1}^M F_m\right), \quad (16)$$

where $\mathbf{S}_k$ is the covariance matrix of $\mathbf{X}_k$ for $k = 1, 2$.

*Proof:* First, the capacity region (15) follows from [11]. Second, using Hadamard's inequality [12], we see that diagonal matrices $\mathbf{S}_1$ and $\mathbf{S}_2$ suffice to achieve all the rate pairs in the capacity region (15), from which we get (16). ∎

*Corollary 1:* When we use each sub-channel instance independently in the strong TPGIC, the following is the sum capacity.

$$\sum_{m=1}^M \min\left(A_m + D_m,\ E_m,\ F_m\right). \quad (17)$$

*Proof:* It follows from *Lemma* 2. ∎

*Lemma 3:* In the low signal-to-noise ratio (SNR) regime, the sum capacity for strong TPGIC under joint coding coincides asymptotically with that under independent coding, i.e.,

$$\lim_{\substack{\max_{k,m} P_{km} \to 0}} \frac{\min\left(\sum_{m=1}^M A_m + D_m,\ \sum_{m=1}^M E_m,\ \sum_{m=1}^M F_m\right)}{\sum_{m=1}^M \min\left(A_m + D_m,\ E_m, F_m\right)} = 1. \quad (18)$$

*Proof:* We refer readers to [16]. ∎

*2) Mixed TPGIC:*

*Lemma 4:* For mixed TPGIC with $\mathbf{H}_{12}^2 \geq \mathbf{H}_{11}^2$ and $\mathbf{H}_{21}^2 \leq \mathbf{H}_{22}^2$, the sum capacity is given by

$$\min\left(\sum_{m=1}^M F_m,\ \sum_{m=1}^M D_m + G_m\right). \quad (19)$$

*Proof:* The sum capacity for the mixed TPGIC is given by

$$\max_{\mathrm{diag}(\mathbf{S}_k) \leq \mathbf{P}_k,\ k=1,\ 2} \min \left\{
\begin{array}{l}
\frac{1}{2}\log_2 \left|\mathbf{I}_M + \mathbf{H}_{12}\mathbf{S}_1\mathbf{H}_{12}^T + \mathbf{H}_{22}\mathbf{S}_2\mathbf{H}_{22}^T\right|, \\
\frac{1}{2}\log_2 \left|\mathbf{I}_M + \mathbf{H}_{11}\mathbf{S}_1\mathbf{H}_{11}^T\left(\mathbf{I}_M + \mathbf{H}_{21}\mathbf{S}_2\mathbf{H}_{21}^T\right)^{-1}\right| \\
\quad + \frac{1}{2}\log_2 \left|\mathbf{I}_M + \mathbf{H}_{22}\mathbf{S}_2\mathbf{H}_{22}^T\right|
\end{array} \right\}, \quad (20)$$

which follows from [11]. Note that the following condition

$$\frac{|\mathbf{I}_M + \mathbf{H}_{22}^2 \mathbf{S}_2|}{|\mathbf{I}_M + \mathbf{H}_{21}^2 \mathbf{S}_2|} \leq \frac{|\mathbf{I}_M + \mathbf{H}_{22}^2 \mathbf{P}_2|}{|\mathbf{I}_M + \mathbf{H}_{21}^2 \mathbf{P}_2|} \quad (21)$$

for any $\mathbf{S}_2$ with $\mathbf{P}_2 = \text{diag}(\mathbf{S}_2)$ is sufficient to show that diagonal covariance matrices $\mathbf{P}_1$ and $\mathbf{P}_2$ are optimal for the sum capacity. Since (21) is satisfied whenever $\mathbf{H}_{21}^2 \leq \mathbf{H}_{22}^2$[1], (20) becomes (19), which completes the proof. ∎

*Remark 1:* Maximizing power for each sub-channel is optimal for (20).

*Corollary 2:* For mixed TPGICs with $\mathbf{H}_{12}^2 \geq \mathbf{H}_{11}^2$ and $\mathbf{H}_{21}^2 \leq \mathbf{H}_{22}^2$, the sum capacity under independent coding is given by

$$\sum_{m=1}^{M} \min(F_m, \ D_m + G_m). \quad (22)$$

*Proof:* It follows from *Lemma* 4. ∎

*Lemma 5:* In the low SNR regime, for mixed TPGICs with $\mathbf{H}_{12}^2 \geq \mathbf{H}_{11}^2$ and $\mathbf{H}_{21}^2 \leq \mathbf{H}_{22}^2$, the sum capacity under joint coding coincides asymptotically with that under independent coding, i.e.,

$$\lim_{\substack{\max_{k,\ m} P_{km} \to 0}} \frac{\min\left(\sum_{m=1}^{M} F_m, \ \sum_{m=1}^{M} D_m + G_m\right)}{\sum_{m=1}^{M} \min(F_m, \ D_m + G_m)} = 1. \quad (23)$$

*Proof:* We refer readers to [16]. ∎

*3) Noisy-interference TPGIC:*

*Lemma 6:* There exist TPGICs that are separable when the channel realization $\mathbf{H}^{[m]}$ at the $m$-th sub-channel is included in $\mathfrak{N}_m = \left\{\mathbf{H}^{[m]} \Big| \frac{|h_{21,m}|}{|h_{22,m}|} + \frac{|h_{12,m}|}{|h_{11,m}|} \leq 1\right\}$ for all $m = 1, 2, \cdots, M$.

*Proof:* It follows from [9], where the separability is proven when the power constraints satisfy a certain condition. ∎

*Remark 2:* For the TPGIC where the channel realization $\mathbf{H}^{[m]}$ at the $m$-th sub-channel satisfies

$$\frac{|h_{21,m}|}{|h_{22,m}|} + \frac{|h_{12,m}|}{|h_{11,m}|} > 1, \ \frac{|h_{21,m}|}{|h_{22,m}|} \leq 1, \text{ and } \frac{|h_{12,m}|}{|h_{11,m}|} \leq 1$$

for all $m = 1, 2, \cdots, M$, the sum capacity is not known even for the TPGIC under independent coding.

*B. Separability*

*1) Strong TPGIC:*

*Theorem 1:* The strong TPGIC is separable iff

$$\mathbf{H}^{[m]} \in \mathfrak{S}_{1m} \text{ for all } m = 1, 2, \cdots, M, \quad (24)$$

$$\mathbf{H}^{[m]} \in \mathfrak{S}_{2m} \text{ for all } m = 1, 2, \cdots, M, \quad (25)$$

or

$$\mathbf{H}^{[m]} \in \mathfrak{S}_{3m} \text{ for all } m = 1, 2, \cdots, M, \quad (26)$$

---
[1]We omit this proof due to space limitations. We refer readers to [16] for the proof of this.

where

$$\mathfrak{S}_{1m} = \left\{\mathbf{H}^{[m]} \Big| 1 + |h_{11,m}|^2 P_{1m} \leq \frac{|h_{21,m}|^2}{|h_{22,m}|^2}, \right.$$
$$\left. 1 + |h_{22,m}|^2 P_{2m} \leq \frac{|h_{12,m}|^2}{|h_{11,m}|^2}\right\},$$

$$\mathfrak{S}_{2m} = \left\{\mathbf{H}^{[m]} \Big| 1 + |h_{11,m}|^2 P_{1m} > \frac{|h_{21,m}|^2}{|h_{22,m}|^2} \geq 1, \right.$$
$$\left. \frac{|h_{12,m}|^2}{|h_{11,m}|^2} \geq \frac{|h_{22,m}|^2 P_{2m}}{|h_{11,m}|^2 P_{1m}} \cdot \left(\frac{|h_{21,m}|^2}{|h_{22,m}|^2} - 1\right) + 1\right\},$$

and

$$\mathfrak{S}_{3m} = \left\{\mathbf{H}^{[m]} \Big| 1 + |h_{22,m}|^2 P_{2m} > \frac{|h_{12,m}|^2}{|h_{11,m}|^2} \geq 1, \right.$$
$$\left. \frac{|h_{12,m}|^2}{|h_{11,m}|^2} < \frac{|h_{22,m}|^2 P_{2m}}{|h_{11,m}|^2 P_{1m}} \cdot \left(\frac{|h_{21,m}|^2}{|h_{22,m}|^2} - 1\right) + 1\right\}.$$

*Proof:* Since (16) is always greater than or equal to (17), we only need to consider the condition for equality in the following:

$$\min\left(\sum_{m=1}^{M} A_m + D_m, \ \sum_{m=1}^{M} E_m, \ \sum_{m=1}^{M} F_m\right)$$
$$\geq \sum_{m=1}^{M} \min\{A_m + D_m, \ E_m, \ F_m\}. \quad (27)$$

A necessary and sufficient condition for the equality is given by

$$A_m + D_m \leq \min\{E_m, \ F_m\}, \text{ for all } m = 1, 2, \cdots, M,$$
$$E_m \leq \min\{A_m + D_m, \ F_m\}, \text{ for all } m = 1, 2, \cdots, M,$$

or

$$F_m \leq \min\{A_m + D_m, \ E_m\} \text{ for all } m = 1, 2, \cdots, M,$$

which completes the proof. ∎

*2) Mixed TPGIC:*

*Theorem 2:* The mixed TPGIC with $\mathbf{H}_{12}^2 \geq \mathbf{H}_{11}^2$ and $\mathbf{H}_{21}^2 \leq \mathbf{H}_{22}^2$ is separable iff

$$\mathbf{H}^{[m]} \in \mathfrak{M}_{1m}^{[1]} \text{ for all } m = 1, 2, \cdots, M \quad (28)$$

or

$$\mathbf{H}^{[m]} \in \mathfrak{M}_{2m}^{[1]} \text{ for all } m = 1, 2, \cdots, M, \quad (29)$$

where

$$\mathfrak{M}_{1m}^{[1]} = \left\{\mathbf{H}^{[m]} \Big| \frac{|h_{12,m}|^2}{|h_{11,m}|^2} \leq \frac{1 + |h_{22,m}|^2 P_{2,m}}{1 + |h_{21,m}|^2 P_{2,m}}\right\} \quad (30)$$

and

$$\mathfrak{M}_{2m}^{[1]} = \left\{\mathbf{H}^{[m]} \Big| \frac{|h_{12,m}|^2}{|h_{11,m}|^2} > \frac{1 + |h_{22,m}|^2 P_{2,m}}{1 + |h_{21,m}|^2 P_{2,m}}\right\}. \quad (31)$$

*Proof:* The sum capacity is given by

$$\min\left(\sum_{m=1}^{M} F_m, \ \sum_{m=1}^{M} D_m + G_m\right)$$

from *Lemma* 4. Under the independent coding, the sum capacity is given by

$$\sum_{m=1}^{M} \min(F_m, \ D_m + G_m)$$

from *Corollary* 2. Note that

$$\min\left(\sum_{m=1}^{M} F_m, \ \sum_{m=1}^{M} D_m + G_m\right) \geq \sum_{m=1}^{M} \min(F_m, \ D_m + G_m),$$

where the necessary and sufficient condition for the equality is given by

$$F_m \leq D_m + G_m \text{ for all } m = 1, 2, \cdots, M \quad (32)$$

or

$$F_m > D_m + G_m \text{ for all } m = 1, 2, \cdots, M, \quad (33)$$

which completes the proof. ∎

*Remark 3:* In the low SNR regime, strong TPGICs are separable asymptotically in the sense of the sum capacity, which follows from *Lemma* 3. Also, in case of mixed TPGICs with $\mathbf{H}_{12}^2 \geq \mathbf{H}_{11}^2$ and $\mathbf{H}_{21}^2 \leq \mathbf{H}_{22}^2$, the separability holds asymptotically, which is confirmed by *Lemma* 5.

*3) Weak TPGIC:* From *Lemma* 6, noisy-interference TPGICs are separable in the sense of the sum capacity. Specifically, single-user decoding at each receiver per sub-channel is enough to achieve the sum capacity. Except for the noisy-interference TPGICs, it is not known if weak TPGICs with $\mathbf{H}_{12}^2 \leq \mathbf{H}_{11}^2$ and $\mathbf{H}_{21}^2 \leq \mathbf{H}_{22}^2$ are separable or not. However, we can conclude the separability partially based on some known inner and outer bounds.

First, for the TPGIC, the sum capacity under independent coding is upper bounded by

$$\sum_{m=1}^{M} \min(A_m + H_m, \ D_m + G_m, \ I_m + J_m) \quad (34)$$

based on the outer bound results in [13] and [14]. Second, for the TPGIC, the sum capacity under joint coding is lower bounded by

$$\max_{\substack{\text{diag}(\mathbf{S}_{kc}) \leq \beta_k \mathbf{P}_k, \ \text{diag}(\mathbf{S}_{kp}) \leq (1-\beta_k)\mathbf{P}_k \ 0 \leq \beta_k \leq 1, \ k=1, 2}} \min(R_{1c} + R_{2c}, \ R_{12c}) + R_{1p} + R_{2p} \quad (35)$$

from the superposition coding based achievable scheme in [15], where

$$R_{1c} = \min\left(\frac{1}{2}\log_2\left|\mathbf{I}_M + \mathbf{H}_{11}\mathbf{S}_{1c}\mathbf{H}_{11}^T\mathbf{Z}_1^{-1}\right|,\right.$$
$$\left.\frac{1}{2}\log_2\left|\mathbf{I}_M + \mathbf{H}_{12}\mathbf{S}_{1c}\mathbf{H}_{12}^T\mathbf{Z}_2^{-1}\right|\right),$$

$$R_{2c} = \min\left(\frac{1}{2}\log_2\left|\mathbf{I}_M + \mathbf{H}_{21}\mathbf{S}_{2c}\mathbf{H}_{21}^T\mathbf{Z}_1^{-1}\right|,\right.$$
$$\left.\frac{1}{2}\log_2\left|\mathbf{I}_M + \mathbf{H}_{22}\mathbf{S}_{2c}\mathbf{H}_{22}^T\mathbf{Z}_2^{-1}\right|\right),$$

$$R_{12c} = \min\left(\frac{1}{2}\log_2\left|\mathbf{I}_M + \left(\mathbf{H}_{11}\mathbf{S}_{1c}\mathbf{H}_{11}^T + \mathbf{H}_{21}\mathbf{S}_{2c}\mathbf{H}_{21}^T\right)\mathbf{Z}_1^{-1}\right|,\right.$$
$$\left.\frac{1}{2}\log_2\left|\mathbf{I}_M + \left(\mathbf{H}_{12}\mathbf{S}_{1c}\mathbf{H}_{12}^T + \mathbf{H}_{22}\mathbf{S}_{2c}\mathbf{H}_{22}^T\right)\mathbf{Z}_2^{-1}\right|\right),$$

$$R_{1p} = \frac{1}{2}\log_2\left|\mathbf{I}_M + \mathbf{H}_{11}\mathbf{S}_{1p}\mathbf{H}_{11}^T\left(\mathbf{I}_M + \mathbf{H}_{21}\mathbf{S}_{2p}\mathbf{H}_{21}^T\right)^{-1}\right|,$$

and

$$R_{2p} = \frac{1}{2}\log_2\left|\mathbf{I}_M + \mathbf{H}_{22}\mathbf{S}_{2p}\mathbf{H}_{22}^T\left(\mathbf{I}_M + \mathbf{H}_{12}\mathbf{S}_{1p}\mathbf{H}_{12}^T\right)^{-1}\right|,$$

where for $k = 1, 2$, $\mathbf{S}_{kc}$ ($\mathbf{S}_{kp}$) is covariance matrix of $\mathbf{X}_{kc}$ ($\mathbf{X}_{kp}$) which is the transmitted signal vector for common (private) information with $\mathbf{X}_k = \mathbf{X}_{kc} + \mathbf{X}_{kp}$, and $\mathbf{Z}_k$'s are defined as

$$\mathbf{Z}_1 = \mathbf{I}_M + \mathbf{H}_{11}\mathbf{S}_{1p}\mathbf{H}_{11}^T + \mathbf{H}_{21}\mathbf{S}_{2p}\mathbf{H}_{21}^T \quad (36)$$

and

$$\mathbf{Z}_2 = \mathbf{I}_M + \mathbf{H}_{12}\mathbf{S}_{1p}\mathbf{H}_{12}^T + \mathbf{H}_{22}\mathbf{S}_{2p}\mathbf{H}_{22}^T. \quad (37)$$

Since we can show there exist TPGICs such that $(35) > (34)$ is satisfied, it is guaranteed that inseparable TPGICs exist in terms of the sum capacity.

## IV. CONCLUSION

We have considered the separability in the sense of the sum capacity in some TPGICs. Since joint coding is more complicated to implement than independent coding, separability result can help us identify channels for which we can lower complexity without any loss in the sum-rate. We have shown necessary and sufficient conditions for the separability for strong and mixed TPGICs. One interesting observation is that unlike weak OPGICs, independent coding is not always sum-rate optimal for the strong and mixed TPGICs. However, in the low SNR regime, the separability holds asymptotically in the strong and mixed TPGICs.

### ACKNOWLEDGMENT

This work was supported by the IT R&D program of MKE/IITA. [2008-F-004-01, 5G mobile communication systems based on beam division multiple access and relays with group cooperation]

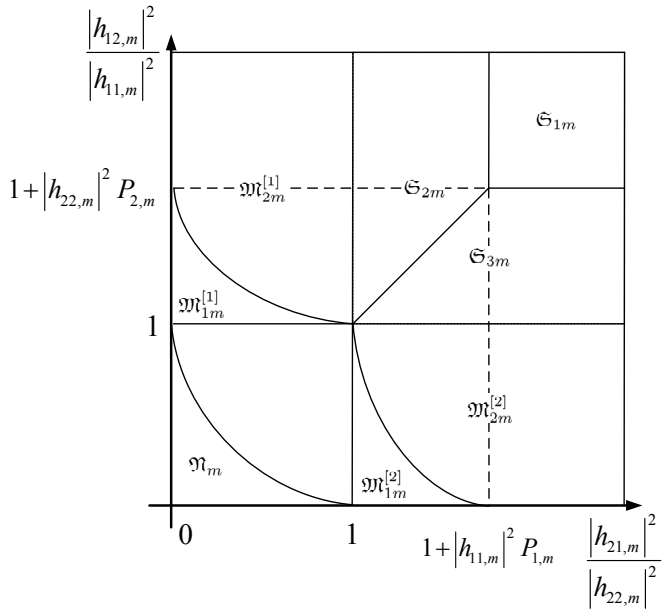

Fig. 1. Separable parallel Gaussian interference channels whose channel realizations are $\mathbf{H}^{[m]} \in \mathfrak{S}_{im}$ for all $m = 1, 2, \cdots, M$, $\mathbf{H}^{[m]} \in \mathfrak{M}_{km}^{[j]}$ for all $m = 1, 2, \cdots, M$, or $\mathbf{H}^{[m]} \in \mathfrak{N}_m$ for all $m = 1, 2, \cdots, M$, where $i = 1, 2, 3$, $j = 1, 2$, and $k = 1, 2$.